\documentclass[a4paper,12 pt]{article} 
\usepackage{camready,epsf,psfig}
\title{ASTROPHYSICAL NEUTRON CAPTURE RATES IN s--
AND r--PROCESS NUCLEOSYNTHESIS} 
\author{H. Beer}
\address{Forschungszentrum Karlsruhe, Institut f\"ur Kernphysik III,\\P.O.
Box 3640, D--760216 Karlsruhe, Germany} 
\author{P. Mohr, H. Oberhummer}
\address{Institut f\"ur Kernphysik, TU Wien,\\ Wiedner Hauptstra{\ss}e
8--10, A--1040 Vienna, Austria} 
\author{T. Rauscher} 
\address{Institut f\"ur Physik, Universit\"at Basel,\\
Klingelbergstra{\ss}e 82, CH--4056
Basel, Switzerland} 
\author{P. Mutti, F. Corvi} \address{CEC,
JRC,Institute for Reference Materials and Measurements,\\ Retieseweg,
B--2440 Geel, Belgium} 
\author{P. V. Sedyshev, Yu. P. Popov}
\address{Frank Laboratory of Neutron Physics, JINR, \\ 141980 Dubna,
Moscow Region, Russia} 
\begin{document} 
\maketitle 
\abstracts{\noindent\textbf{Abstract}: 
Astrophysical neutron capture rates of light and heavy nuclei have been
measured and calculated.  For the measurements the activation technique
was applied at the 3.75 MV Karlsruhe Van de Graaff accelerator, and at the
Geel electron linear accelerator (GELINA)  the time--of--flight (TOF)
method was used.  The calculations were performed using direct and
compound nuclear capture models.} 
\section{Introduction} 
The formation of the chemical elements in the astrophysical s-- and
r--process scenarios is controlled by astrophysical neutron capture rates
at thermonuclear energies in the range at 5 to 250\,keV. At these low
projectile energies the reaction energy in the capture process mainly
comes from the neutron binding energy. As, in general, heavy nuclei have a
high density of excited states at this energy the well--known compound
nuclear capture reaction mechanism (CN) is dominant. But in light,
especially, neutron--rich isotopes, and in isotopes at or close to magic
neutron and/or proton shells only few excited states are expected. In
these cases besides the compound capture another another capture
mechanism, direct capture (DC), is comparable or even
dominant.\,\cite{ohu96}$^-$\,\cite{bcm97}
In the measurements activation techniques and the time--of--flight method
were employed. The measurements were analyzed and the results compared
with compound and direct capture model calculations. Cross sections of
unstable target nuclei were estimated theoretically. The astrophysical
capture rates were applied in s--process calculations.  
\section{Activation Measurements} 
\subsection{Activation Method} 
\begin{figure}[t]
\begin{minipage}[t]{95mm} 
\psfig{figure=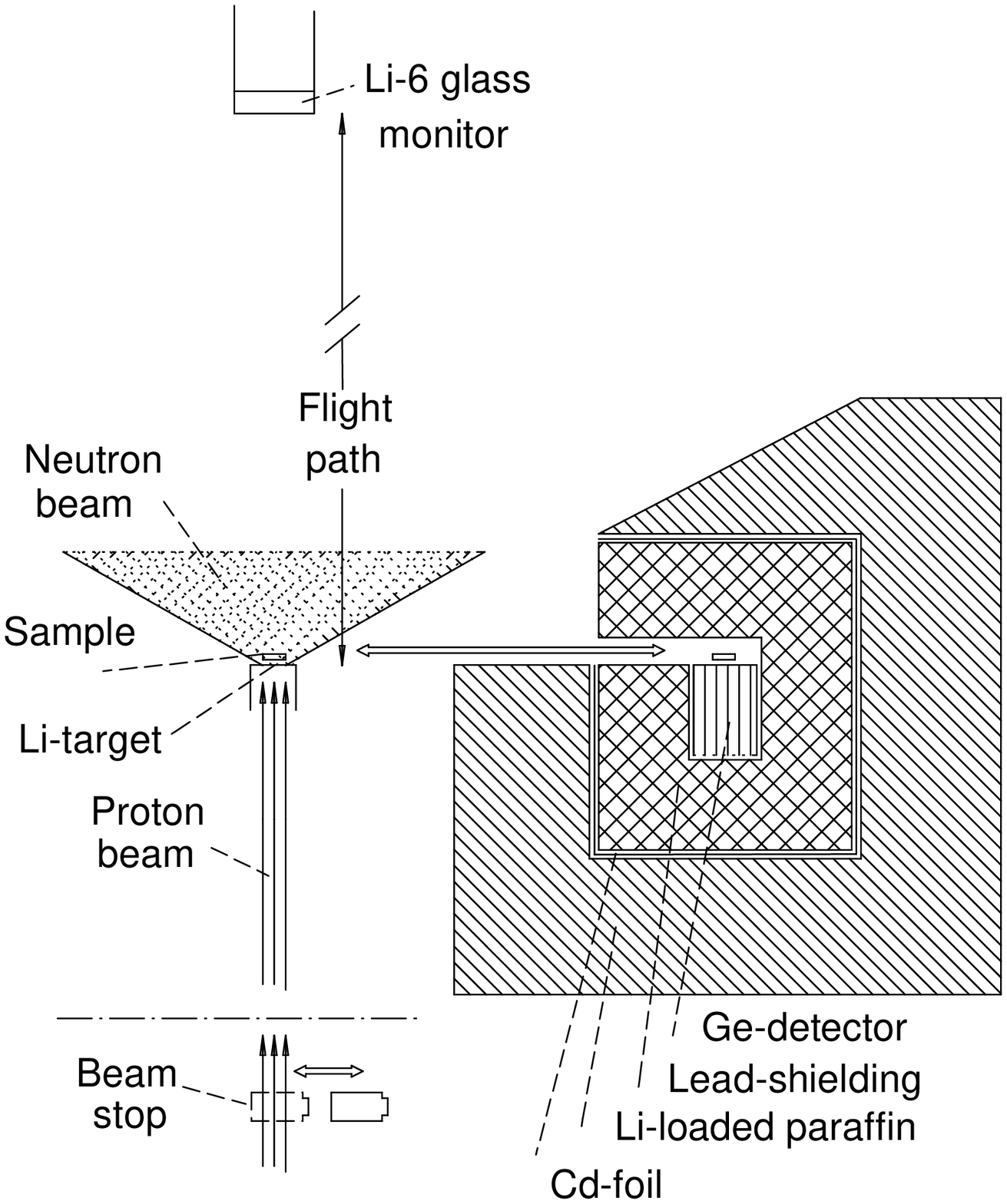,width=95mm}
\caption{\label{setup} Setup of the fast cyclic activation technique.}
\end{minipage} 
\hspace{\fill} 
\begin{minipage}[t]{60mm}
\psfig{figure=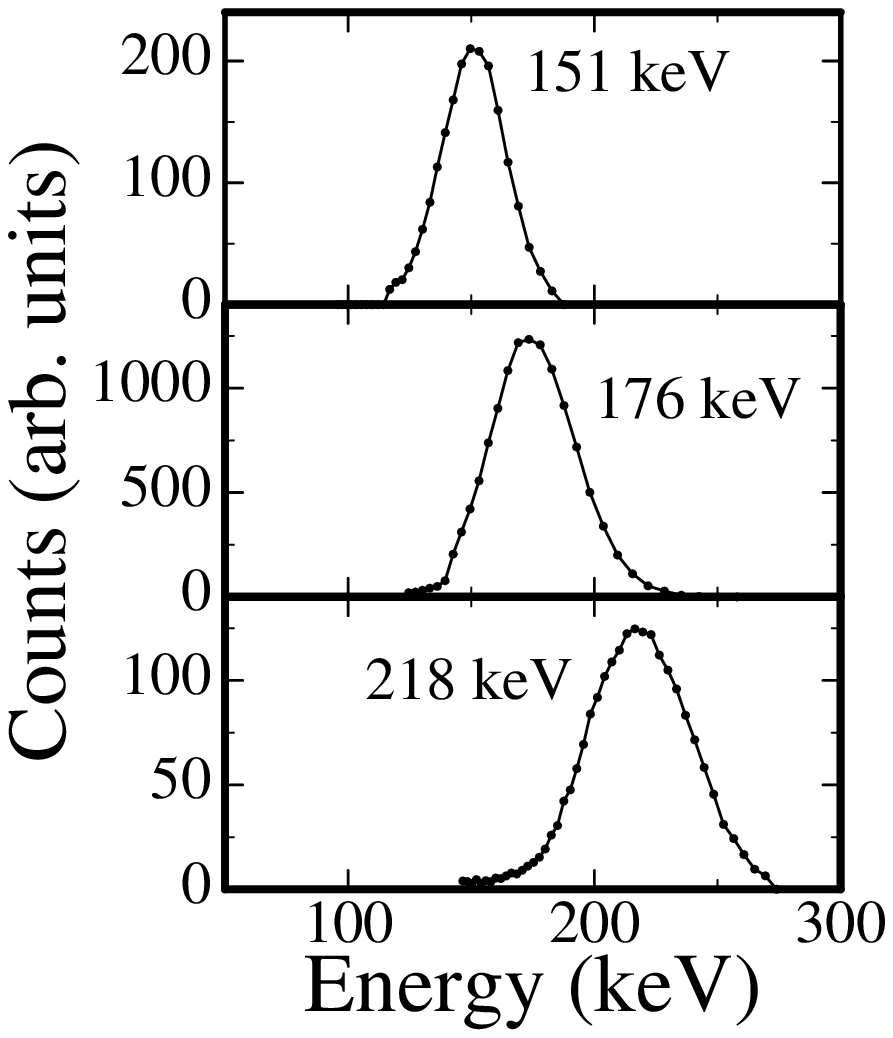,width=60mm} 
\caption{\label{neuspek} Neutron spectra used for
activation.} 
\end{minipage}

\end{figure}

In the activation method
a sample of the target isotope is irradiated by a neutron spectrum of a
well--defined energy distribution, and the cross section is determined
from the activity of the residual nuclei.  Because of the short
half--lives of many product isotopes a special activation
method\,\cite{bee80,be94} had to be developed at the Karlsruhe 3.75\,MV
Van de Graaff accelerator, the cyclic activation technique, where the
irradiation and activity counting phases which form an activation cycle
are repeated many times (Fig.\,\ref{setup}).  
In the cyclic activation method the time constants for each cycle adjusted
to the decay rate $\lambda$ of the investigated isotope are the irradiation
time $t_{\rm b}$, the counting time
$t_{\rm c}$, the
waiting time $t_{\rm w}$ (the time to switch from the
irradiation to the counting phase),
and the total
time $T=t_{\rm b}+t_{\rm w}+t_{\rm c}+t'_{\rm w}$ ($t'_{\rm w}$ the time
to switch from the
counting to the irradiation phase).
The accumulated number of counts from a total of $n$ cycles,
$C=\sum_{i=1}^n C_i$, with the $C_i$,
the counts of the i--th cycle, irradiated by a neutron flux $\Phi_i$,
is given by
\begin{figure}[t]
\centerline{\psfig{figure=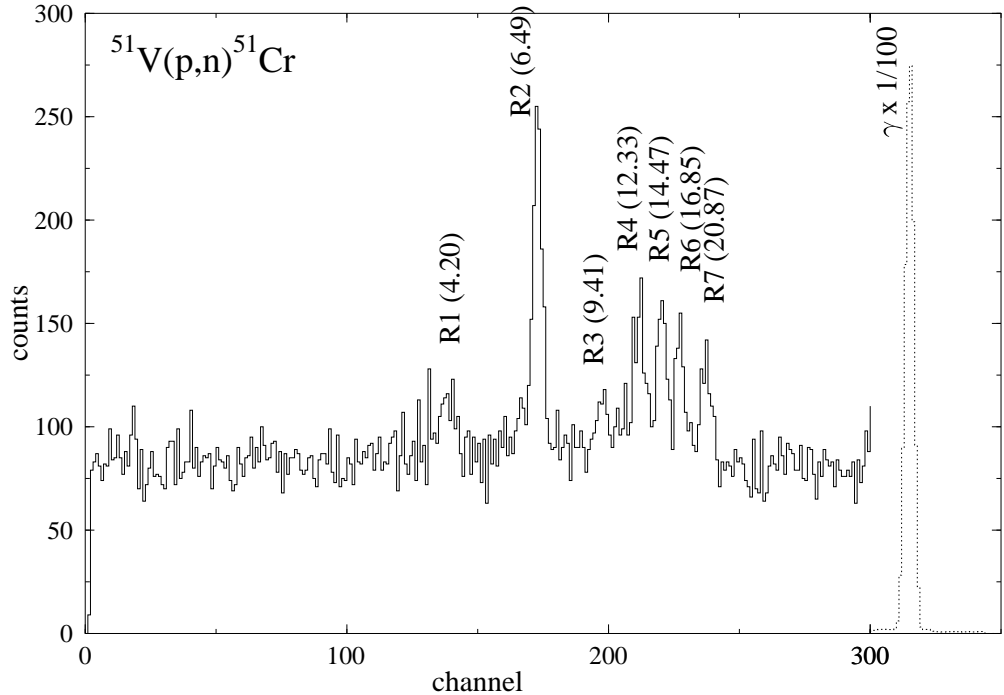,width=145mm}}
\caption{\label{v51} TOF-spectrum of the neutrons from the $^{51}$V(p,n)
reaction close to the reaction threshold.}
\end{figure}
\begin{eqnarray}
\label{eq1}
C=\epsilon_{\gamma}K_{\gamma}f_{\gamma}\lambda^{-1}[1-\exp(-\lambda t_c)]
\exp(-\lambda t_w)
 \frac{1-\exp(-\lambda t_b)}{1-\exp(-\lambda T)} N \sigma
[1-f'_b \exp(-\lambda T)]
 \sum_{i=1}^n \Phi_i  \\
\mbox{with}\qquad
f'_b=\sum_{i=1}^n \Phi_i \exp[-(n-i)\lambda
T] /\sum_{i=1}^n \Phi_i dt\nonumber
\quad .
\end{eqnarray}
The following quantities have been defined: $\epsilon_\gamma$:
Ge--efficiency, $K_\gamma$:
$\gamma$-ray absorption, $f_\gamma$: $\gamma$-ray intensity per decay,
$N$: the number of target
nuclei, $\sigma$: the capture cross section.
The activities of nuclides with half--lives of several hours to days can also be
counted after
the end of the cyclic activation consisting of $n$ cycles:
\begin{eqnarray}
\label{eq3}
C_n=\epsilon_\gamma K_\gamma f_\gamma \lambda^{-1} [1-\exp(-\lambda
T_M)]\exp(-\lambda T_W)
[1-\exp(-\lambda t_b)] N \sigma f'_b \sum_{i=1}^n \Phi_i
\end{eqnarray}
Here $T_{\rm M}$ is the measuring time of the Ge--detector and $T_{\rm W}$
the time elapsed between the
end of cyclic activation and beginning of the new data acquisition.
To avoid the absolute measurement of the neutron flux the sample to be
investigated is sandwiched between two thin gold foils and the capture
cross section $\sigma$ is determined relative to the well--known 
$^{197}$Au cross section.\,\cite{rat88}
\begin{figure}[t]
\begin{minipage}[t]{88mm}
\centerline{\psfig{figure=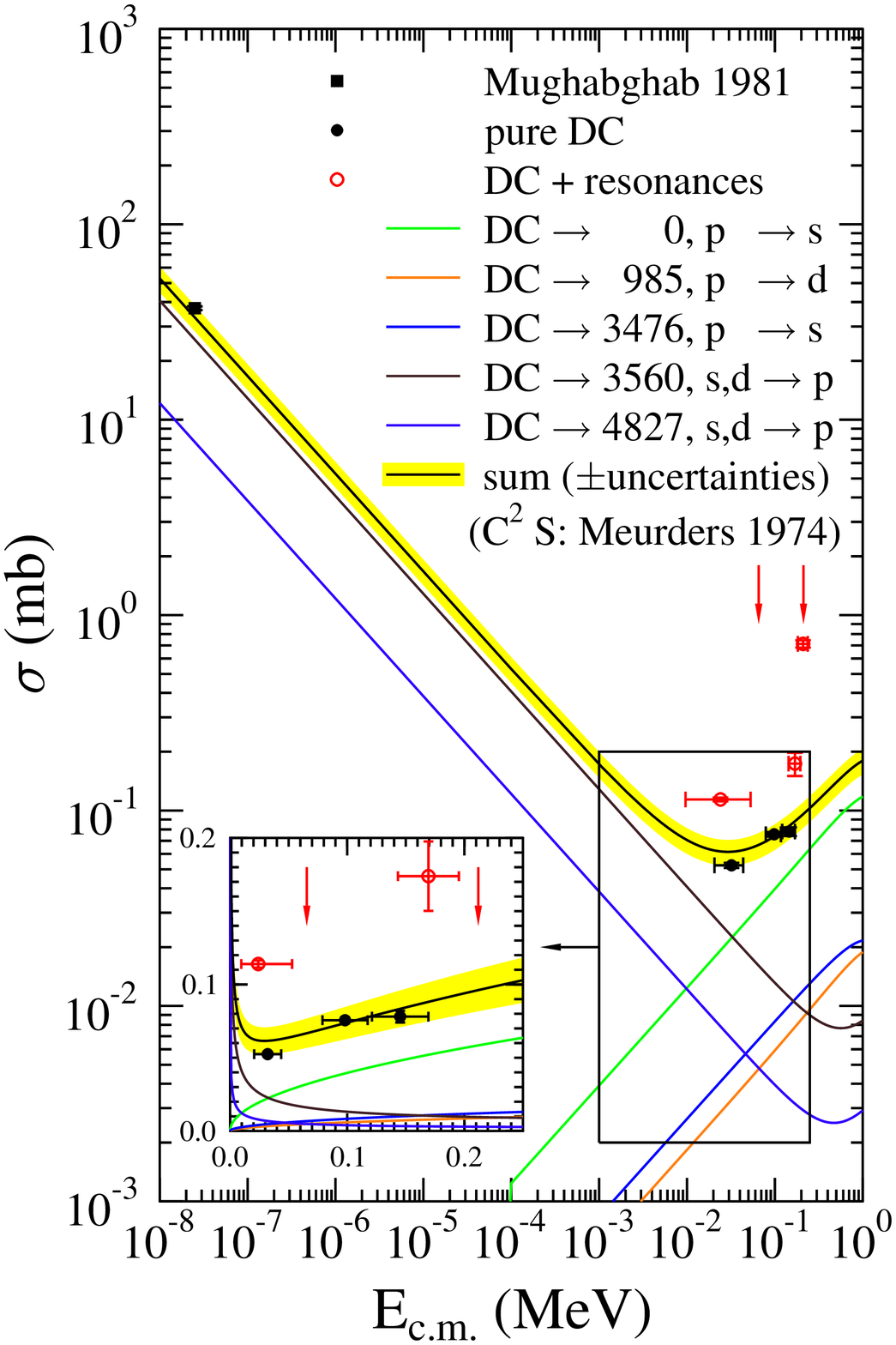,width=88mm}}
\caption{\label{26Mg} Comparison of the DC cross section for
$^{26}$Mg(n,$\gamma$)$^{27}$Mg
with the experimental data from thermal to thermonuclear projectile energies.
The DC contributions for the different transitions to the final states of
$^{27}$Mg as well as the sum of all transitions (solid curve) are shown.}
\end{minipage}
\hspace{\fill}
\begin{minipage}[t]{70mm}
\psfig{figure=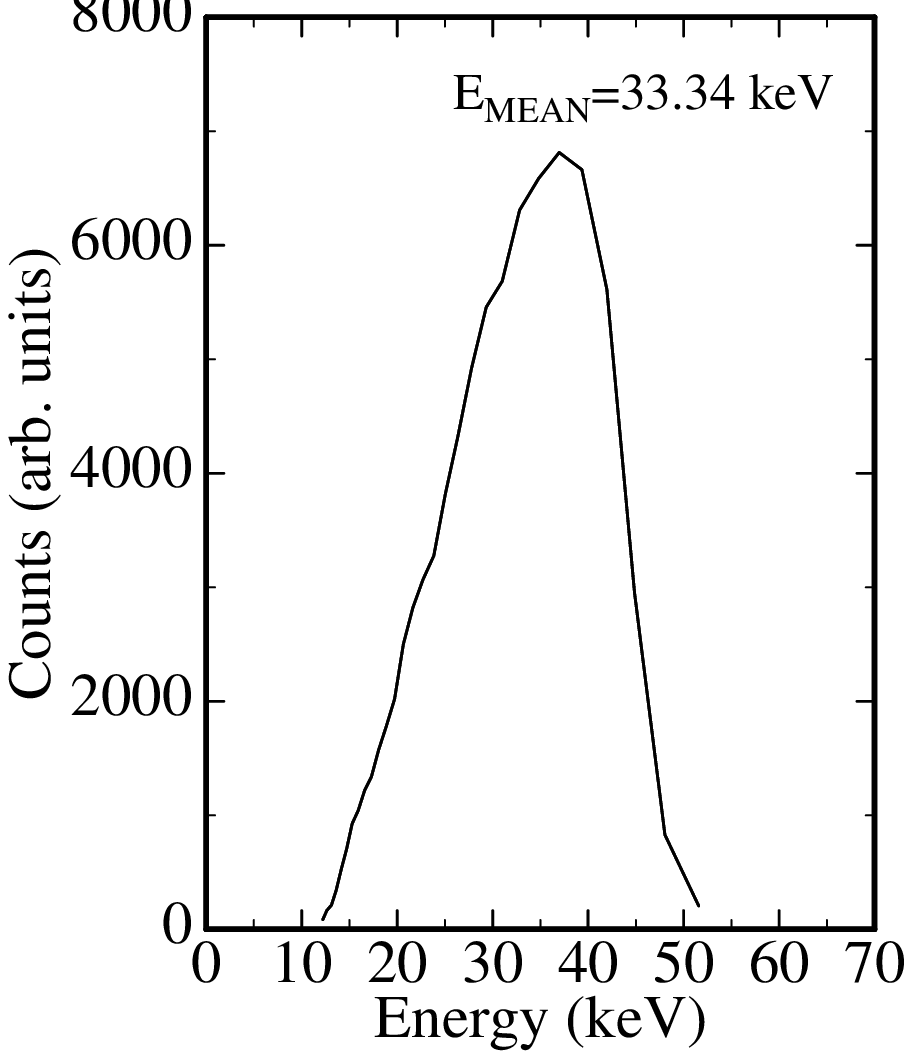,width=70mm}
\caption{\label{spek30} Neutron spectrum at the threshold of the
$^{7}$Li(p,n) reaction used for the measurement of the $^{26}$Mg
cross section at 33\,keV. Note that with these neutrons                 
the resonance at 68.7\,keV is
not excited.}
\end{minipage}
\end{figure}
\begin{figure}[t]
\centerline{\psfig{figure=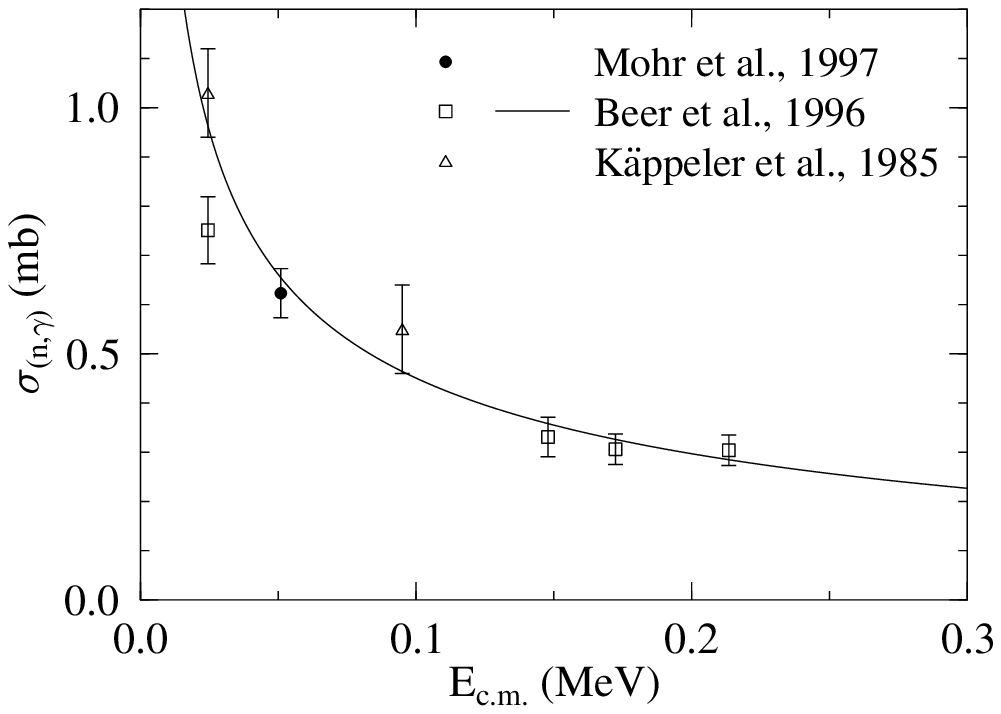,width=100mm}}
\caption{\label{48Ca} Comparison of the DC cross section for
$^{48}$Ca(n,$\gamma$)$^{49}$Ca
with the experimental data at thermonuclear projectile energies.
}
\end{figure}

Normally the neutrons are generated by the $^7$Li(p,n) and T(p,n)
reactions. For the energy points at
25\,keV and 52\,keV one can take advantage of the
special properties of these reactions at the reaction threshold. For
neutron energies above 100\,keV the $^7$Li(p,n) reaction is
applied at higher proton energies and with thin Li--targets
(full half--width 15--20\,keV).
Examples of generated neutron spectra used in the measurements
are shown in Fig.\,\ref{neuspek}. For light isotopes reactor neutrons
at 25.3\,meV can be an important supplement of the measurements. This
has been demonstrated for $^{36}$S\,\cite{bee95} and above all for
$^{48}$Ca.\,\cite{bee96,moh97}
Additionally, it is desirable to use neutrons in the energy range
5 to 12\,keV. In our investigations of (p,n) reactions of different
isotopes at reaction threshold we found two possible candidates.
With the $^{51}$V(p,n) reaction neutrons around
6 keV can be generated. The Fig.\,\ref{v51} shows a time--of--flight spectrum
of the $^{51}$V(p,n) reaction with a prominent resonance at 6.49 keV.
A similar neutron source for neutrons of 8 keV represents the $^{45}$Sc(p,n)
reaction. This reaction is still under investigation.
\subsection{$^{26}$Mg and $^{48}$Ca Measurements}
Astrophysical neutron capture rates of light isotopes are required for
inhomogeneous big bang scenarios, the $\alpha$--rich freeze out, and the
s--process of light isotopes to study neutron
poisons, the s--process production of rare neutron--rich isotopes
($^{36}$S, $^{40}$Ar), and isotopic anomalies (Si, Ti)\,\cite{hop94}.
Neutron capture of light isotopes at the border of $\beta$ stability is
often dominated by the direct capture mechanism due to the lack of
resonance states of the compound system. Direct capture is a smoothly
varying cross section normally smaller than 1\,mbarn in the energy range
above 1\,keV. In the Figs.\,\ref{26Mg},\ref{48Ca} the recent results
of the measurements on $^{26}$Mg\,\cite{moh96} and $^{48}$Ca\,\cite{moh97}
are shown. The data can be described adequately theoretically by the
direct capture process.\,\cite{ohu96,kim87} In the keV energy range
$^{26}$Mg lies in a transition region from s-- to p--wave direct capture.
Therefore, the cross section (Fig.\,\ref{26Mg}) which is decreasing up to
30\,keV is increasing towards higher energies.\,\cite{moh96} The total DC
capture is the sum of many individual transitions. This would make its
measurement via these individual partial transitions complicated. At three
of the six data points the DC capture is superimposed by compound capture
of individual resonances at 68.7 and 220\,keV (arrows in
Fig.\,\ref{26Mg}). The influence of the
68.7\,keV resonance on our value at 33\,keV was excluded by choosing a
neutron spectrum at the threshold of the $^7$Li(p,n) reaction
(Fig.\,\ref{spek30}). In Fig.\,\ref{48Ca} our\,\cite{bee96,moh97} and
previous\,\cite{kaep85} $^{48}$Ca measurements at thermonuclear energies
are summarized. Our new data point at 52\,keV confirms the 1/v behavior
of this cross section which is the result of direct s--wave capture. No
sign of a destructive interference between the non--resonant direct
capture and a hypothetical 1/2$^{+}$ s--wave resonance at 1.5\,keV was
found.\,\cite{bee96}
\subsection{Measurements of Pt--Isotopes}
\begin{figure}[t]
\begin{minipage}[t]{155mm}
\centerline{\psfig{figure=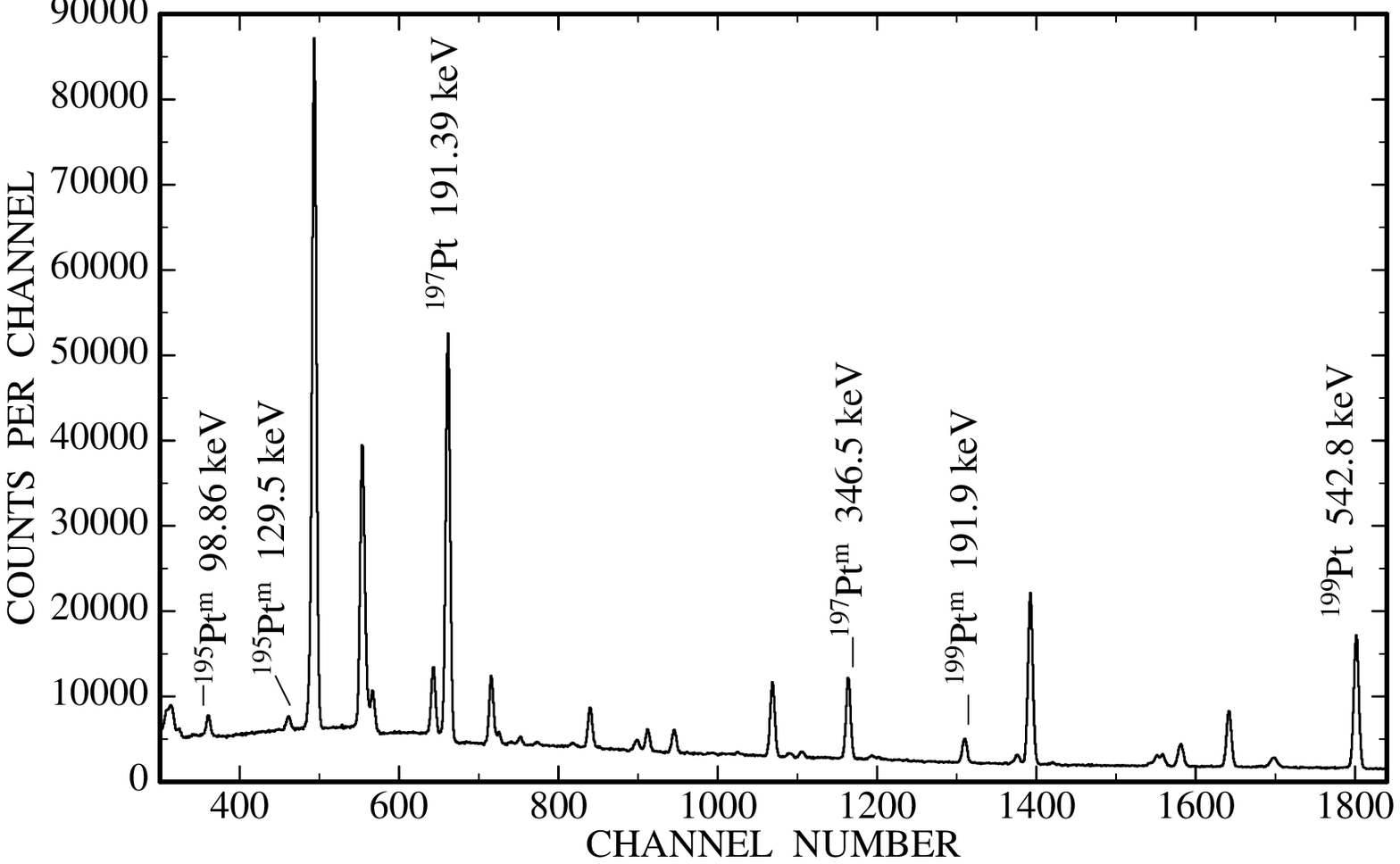,width=155mm}}
\caption{\label{pt1} Accumulated intensities of decay $\gamma$--ray
lines from $^{195m}$Pt, $^{197}$Pt, $^{197m}$Pt, $^{199}$Pt, and
$^{199m}$Pt. Accumulation of 4224 activation cycles.
}
\end{minipage}
\begin{minipage}[b]{155mm}
\centerline{\psfig{figure=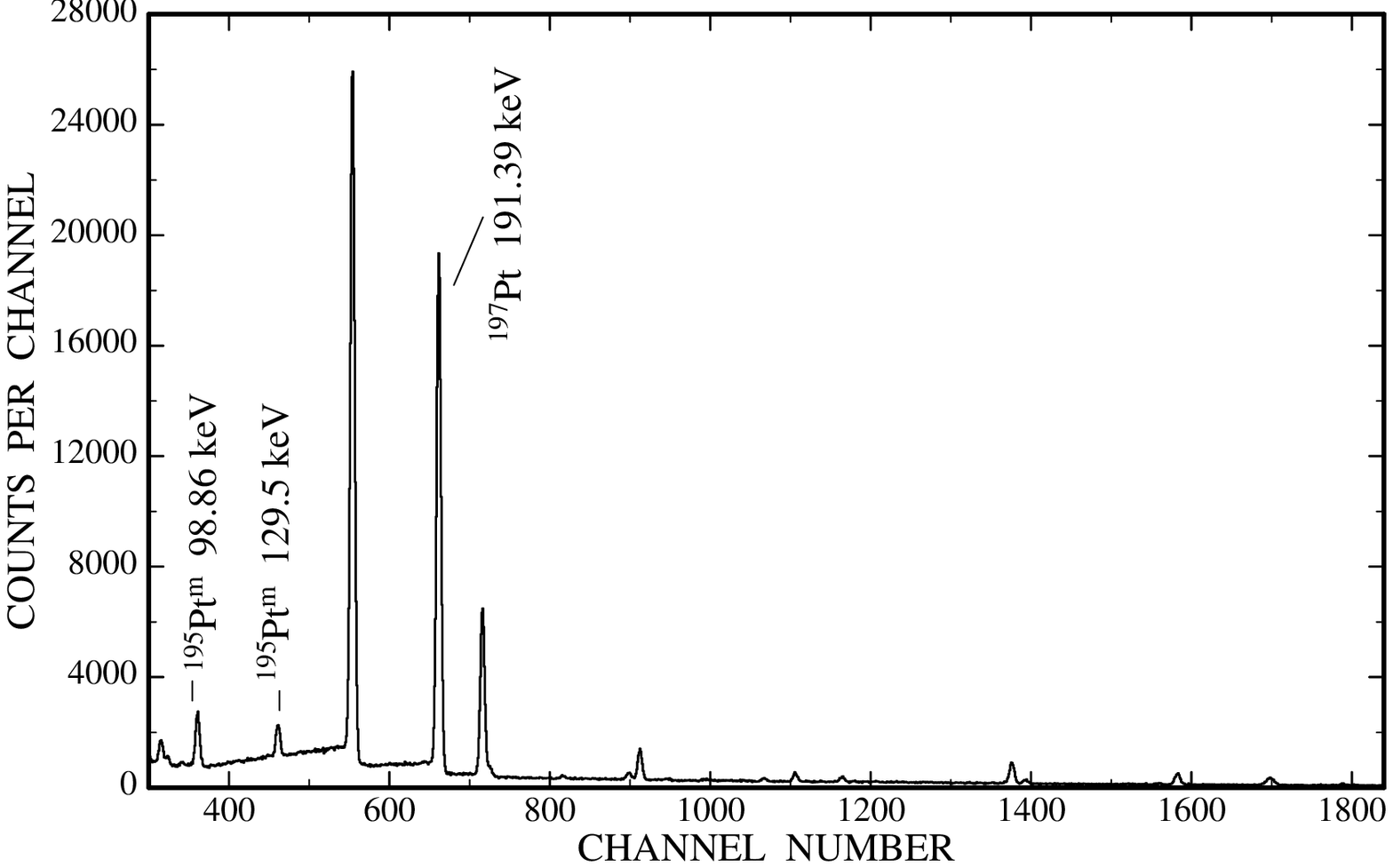,width=155mm}}
\caption{\label{pt2} The intensities of decay $\gamma$--ray lines from
$^{195m}$Pt and $^{197}$Pt. The data acquisition after the termination of
the cyclic activation run. Note that in this case only the long lived
activities can be counted. }
\end{minipage}
\end{figure}

The Pt--isotopes have not yet been investigated by a time--of--flight
experiment. With the activation technique capture measurements of
$^{190,194,196,198}$Pt could be performed with natural Pt samples (metallic
foils of 6 mm diameter) even for the rare $^{190}$Pt isotope
(isotopic content 0.01\,\%). Although the half lives of the generated
activities reach from 13.6\,s to 4\,d all isotopic cross sections could be
measured in one run. The Figs.~\ref{pt1},\ref{pt2} show the
$\gamma$--ray activities of the individual Pt--isotopes accumulated during
and after termination of the cyclic activation measurement. During the
cyclic activation $\gamma$--line intensities of all isotopes also of those
with short lived activities can be observed (Fig.~\ref{pt1}), whereas
after the cyclic activation only the long lived activities are accumulated
(Fig.~\ref{pt2}). This demonstrates that the
cyclic activation technique is an extension of the common activation
method\,\cite{bee80}
and, therefore, applicable to radioactivities of short and long half--lives as
well.\,\cite{be94} It is free
of saturation effects which limit the reasonable irradiation time of
a common activation.
The results of our measurements\,\cite{bt97} are shown in
Fig.\,\ref{xpt} together with estimates for the isotopes $^{192,195}$Pt.
Additionally statistical model calculations with the code
SMOKER\,\cite{tat87} have been carried out.
In Fig.~\ref{xpt} also a comparison is made between our measured and
theoretically calculated Pt Maxwellian average capture (MAC) cross
sections (BR)\,\cite{bt97} and
previous theoretical results from (AGM)\,\cite{allen71},
(HWFZ)\,\cite{Holmes76}, (Harris)\,\cite{harris81}, and
(CTT)\,\cite{Cow91}. Excellent agreement between experimental and
theoretical values was found for our MAC
cross sections.

The s--process nucleosynthesis path in the Os to Pt mass region is shown
in Fig.~\ref{pathy}. For $^{191}$Os, $^{192}$Ir, and
$^{193}$Pt there is a competition between $\beta$--decay and neutron
capture. The $\beta$--decay half lives are dependent on temperature
and electron density of the s--process environment.\,\cite{tak87}
The abundance of s--only $^{192}$Pt originates from the branchings
at  $^{191}$Os and $^{192}$Ir. The isotopes $^{190}$Pt and
$^{198}$Pt are not on the s--process path, therefore the seed abundances  
vanish during nucleosynthesis.
Calculations to reproduce the s--process abundances in the Os-- to
Pt--mass region have been carried out using parametrized models
described in detail elsewhere.\,\cite{bcm97}
In the analysis of the present branching especially the
neutron density was adjusted to reproduce the solar abundance of
the s--only isotope
$^{192}$Pt. \hfill The \hfill calculations \hfill were \hfill carried
\hfill out \hfill taking \hfill into \hfill account \hfill the \hfill 
uncertainties \hfill of \hfill the
\clearpage
\clearpage
\begin{figure}[t]
\centerline{\psfig{figure=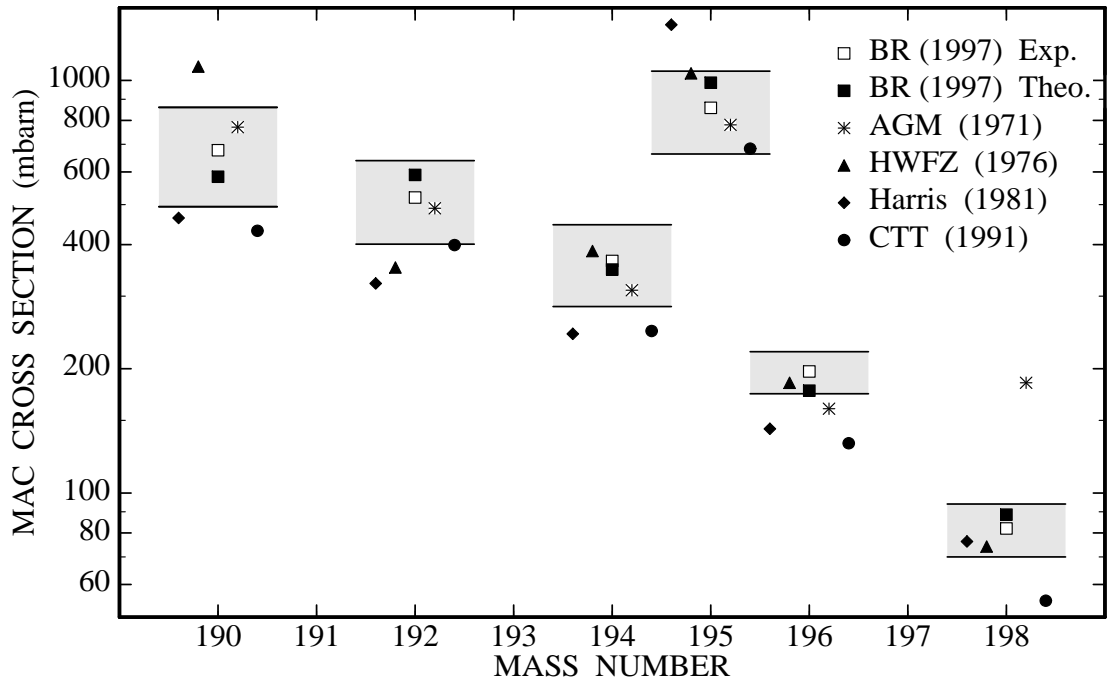,width=150mm}}
\caption{\label{xpt} MAC cross sections of the stable Pt--isotopes at
$kT$=30 keV. Comparison of the experimentally and theoretically determined
values of Beer and Rauscher (BR 1997) with theoretical
estimates from literature. Note that the theoretically calculated values of
BR (1997) agree with the experimental estimates within quoted
uncertainties (shaded area). The references of the different
abbreviations in the figure are given in the text.}
\end{figure}
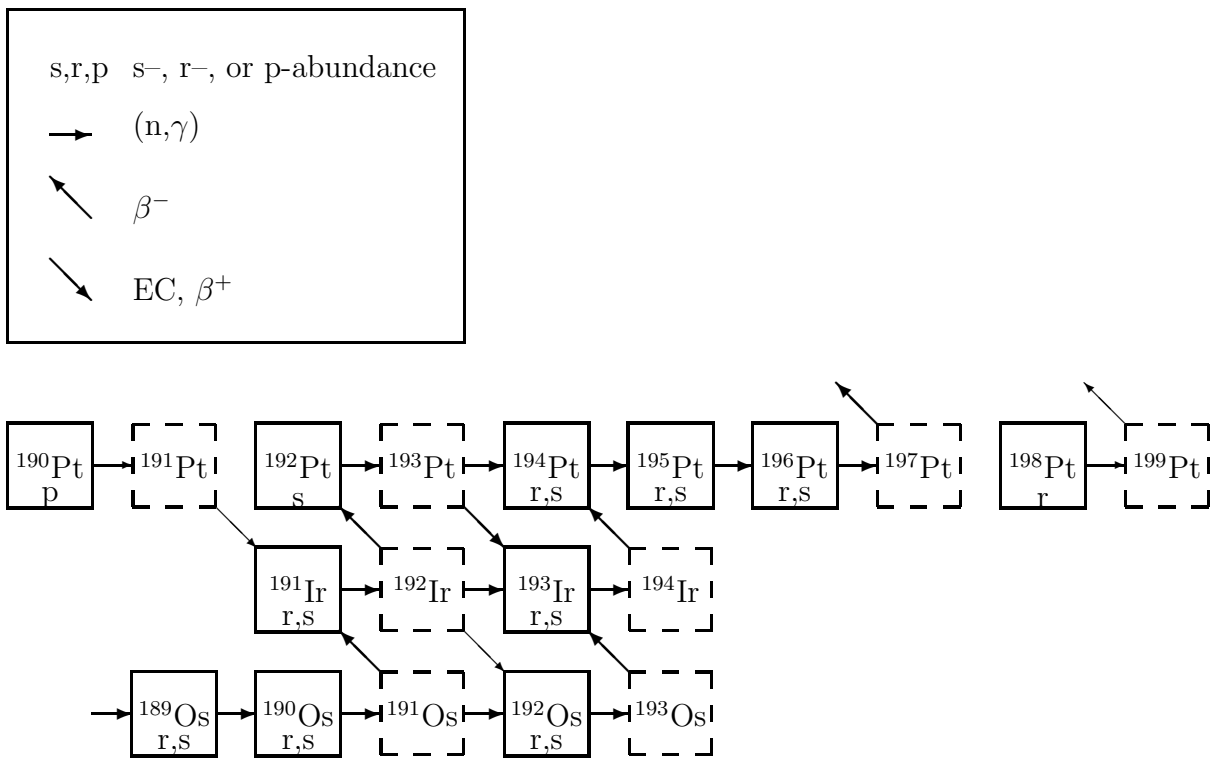
\begin{figure}[b]
\unitlength1.1cm
\begin{picture}(15,10)
\thicklines
\put(0.0,5.0){\framebox(5.5,4.0)}

\put(0.5,8.2){s,r,p}
\put(1.5,8.2){s--, r--, or p-abundance}
\put(0.5,7.5){\vector(1,0){0.5}}
\put(1.5,7.5){(n,$\gamma$)}

\put(1.0,6.5){\vector(-1,1){0.5}}
\put(1.5,6.5){$\beta^-$}

\put(0.5,6.0){\vector(1,-1){0.5}}
\put(1.5,5.5){EC, $\beta^+$}

\put(1.0,0.5){\vector(1,0){0.5}}
\put(1.5,0.0){\framebox(1,1){$^{189}$Os}}
\put(1.5,0.0){\framebox(1,1)[b]{r,s}}
\put(2.5,0.5){\vector(1,0){0.5}}
\put(3.0,0.0){\framebox(1,1){$^{190}$Os}}
\put(3.0,0.0){\framebox(1,1)[b]{r,s}}
\put(4.0,0.5){\vector(1,0){0.5}}
\put(4.5,0.0){\dashbox{0.2}(1,1){$^{191}$Os}}
\put(4.5,1.0){\vector(-1,1){0.5}}
\put(5.5,0.5){\vector(1,0){0.5}}
\put(6.0,0.0){\framebox(1,1){$^{192}$Os}}
\put(6.0,0.0){\framebox(1,1)[b]{r,s}}
\put(7.0,0.5){\vector(1,0){0.5}}
\put(7.5,0.0){\dashbox{0.2}(1,1){$^{193}$Os}}
\put(7.5,1.0){\vector(-1,1){0.5}}

\put(3.0,1.5){\framebox(1,1){$^{191}$Ir}}
\put(3.0,1.5){\framebox(1,1)[b]{r,s}}
\put(4.0,2.0){\vector(1,0){0.5}}
\put(4.5,1.5){\dashbox{0.2}(1,1){$^{192}$Ir}}
\put(5.5,2.0){\vector(1,0){0.5}}
\put(4.5,2.5){\vector(-1,1){0.5}}
\thinlines
\put(5.5,1.5){\vector(1,-1){0.5}}
\thicklines
\put(6.0,1.5){\framebox(1,1){$^{193}$Ir}}
\put(6.0,1.5){\framebox(1,1)[b]{r,s}}
\put(7.0,2.0){\vector(1,0){0.5}}
\put(7.5,1.5){\dashbox{0.2}(1,1){$^{194}$Ir}}
\put(7.5,2.5){\vector(-1,1){0.5}}

\put(0.0,3.0){\framebox(1,1){$^{190}$Pt}}
\put(0.0,3.0){\framebox(1,1)[b]{p}}
\thinlines
\put(1.0,3.5){\vector(1,0){0.5}}
\thicklines
\put(1.5,3.0){\dashbox{0.2}(1,1){$^{191}$Pt}}
\thinlines
\put(2.5,3.0){\vector(1,-1){0.5}}
\thicklines
\put(3.0,3.0){\framebox(1,1){$^{192}$Pt}}
\put(3.0,3.0){\framebox(1,1)[b]{s}}
\put(4.0,3.5){\vector(1,0){0.5}}
\put(4.5,3.0){\dashbox{0.2}(1,1){$^{193}$Pt}}
\put(5.5,3.5){\vector(1,0){0.5}}
\put(5.5,3.0){\vector(1,-1){0.5}}
\put(6.0,3.0){\framebox(1,1){$^{194}$Pt}}
\put(6.0,3.0){\framebox(1,1)[b]{r,s}}
\put(7.0,3.5){\vector(1,0){0.5}}
\put(7.5,3.0){\framebox(1,1){$^{195}$Pt}}
\put(7.5,3.0){\framebox(1,1)[b]{r,s}}
\put(8.5,3.5){\vector(1,0){0.5}}
\put(9.0,3.0){\framebox(1,1){$^{196}$Pt}}
\put(9.0,3.0){\framebox(1,1)[b]{r,s}}
\put(10.0,3.5){\vector(1,0){0.5}}
\put(10.5,3.0){\dashbox{0.2}(1,1){$^{197}$Pt}}
\put(10.5,4.0){\vector(-1,1){0.5}}
\put(12.0,3.0){\framebox(1,1){$^{198}$Pt}}
\put(12.0,3.0){\framebox(1,1)[b]{r}}
\thinlines
\put(13.0,3.5){\vector(1,0){0.5}}
\thicklines
\put(13.5,3.0){\dashbox{0.2}(1,1){$^{199}$Pt}}
\thinlines
\put(13.5,4.0){\vector(-1,1){0.5}}

\end{picture}
\caption{\label{pathy} s--process path between Os and Pt. The solid boxes
are the stable isotopes. The origin of nucleosynthesis is indicated.
If there are two contributions, the major contribution is designated first.
The dashed boxes are radioactive nuclei.
}
\end{figure}
\clearpage
\noindent
MAC cross sections and the solar $^{192}$Pt
abundance. At the neutron density $5.0\times 10^8$\,cm$^{-3}$ which was
deduced using all s--process branchings\,\cite{bcm97} the $^{192}$Pt
abundance appears to be distinctly
underproduced compared with the respective solar abundance.
Only at definitely lower
neutron densities
($\le 3.2\times 10^8$ cm$^{-3}$)
the\hfill calculated abundance
values become consistent
with the range of empirical $^{192}$Pt values. This distinct difference
between neutron densities describing all s--process branchings
on the average and the smaller values found for the $^{191}$Os-- and
$^{192}$Ir--branchings could be the effect of freeze--out for the
$^{22}$Ne($\alpha$,n) neutron source. To confirm this effect the MAC cross
sections of the unstable $^{191}$Os and $^{192}$Ir isotopes have to be
determined to better than 20\,\%.
\section{Time--of--Flight Measurements}
\subsection{$^{136}$Ba and $^{209}$Bi Results}
Among the heavy nuclei the isotopes near or at magic neutron shells have
got the smallest capture cross sections. These bottleneck isotopes which
control the s--process nucleosynthesis are characterized below 100\,keV
neutron energy by resonance structure which can be well resolved in high
resolution time--of--flight measurements. At the electron linear
accelerator GELINA in Geel, Belgium, neutron capture measurements of
the bottleneck isotopes $^{138}$Ba and $^{208}$Pb have been
carried out using a pair of
C$_6$D$_6$ liquid scintillation detectors at the 60\,m flight path
station.\,\cite{bcm97} Details of the detection geometry and the
characteristics of the detector efficiency can be found
elsewhere.\,\cite{cor91}
These experiments were continued with new measurements of
$^{136}$Ba\,\cite{mut96}
and $^{209}$Bi\,\cite{mut97}. Both measurements are in agreement within
quoted uncertainties with previous results.\,\cite{vwk95,ksw96,mug84}
The stellar rate of s--only $^{136}$Ba is now one of the best known
astrophysical rates.\,\cite{mut96,vwk95,ksw96} It should be noted that the
MAC cross section of $^{209}$Bi at $kT$=30\,keV reported by
Mughabghab\,\cite{mug84}
is an erroneous calculation from the resonance parameters.

With these reliable MAC cross sections of the bottleneck isotopes
the s--process calculations were improved.
Parametrized s--process model calculations have been carried
out.\,\cite{bcm97} In the standard SP--model it is attempted
to fit global and local structures of the synthesis by one
major neutron source burning at $kT$=27.1\,keV. A better
reproduction of the solar s--process abundances was obtained
in the DP--model by combined burning of two neutron sources
ignited at $kT$=8 and 28\,keV, respectively. The new MAC cross
section of $^{209}$Bi confirms this s--process analysis at
the termination of the synthesis.\,\cite{bcm97}
\begin{table}[b]
\caption{Ba isotopic ratios. A non--solar s--process composition
can be assumed for s-Ba from meteoritic SiC grains. In brackets the
results expected from a solar s--process composition.}
\label{t2a}
\begin{tabular*}{\textwidth}{@{}l@{\extracolsep{\fill}}cccccc}
\hline
&$\tau_0$&\multicolumn{1}{c}{134/136} &
\multicolumn{1}{c}{135/136} &
\multicolumn{1}{c}{137/136} &
\multicolumn{1}{c}{138/136} \\
&(mbarn$^{-1}$)&&&&\\
\hline
solar ratios$^{\rm a}$&&0.308&0.839&1.430&9.129\\
s--Ba$^{\rm b}$&&0.341$\pm$0.003&0.135$\pm$0.003&0.684$\pm$0.007&
6.22$\pm$0.08\\
SP model& 0.188 & 0.37 & 0.15 & 0.74 & 6.22 \\
        &(0.296)&(0.35)&(0.14)&(0.75)&(7.89)\\
DP model& 0.102 &0.31  & 0.16 & 0.76 & 6.22  \\
        &(0.140)&(0.30)&(0.15)&(0.76)&(7.21)\\
\hline
\end{tabular*}
{\footnotesize $^{\rm a}$ Ref.\cite{pabe93}, $^{\rm b}$ Ref.\cite{pro93}
}
\end{table}
The special role of the $^{138}$Ba MAC cross section and the new
measurement of $^{136}$Ba\,\cite{mut96} also allows a more detailed
investigation of the isotopic anomalies in barium ascribed to the
s--process. In Table\,\ref{t2a} Ba isotopic ratios
obtained from s--process calculations are compared with the observed
s--Ba ratios detected in meteoritic SiC grains\,\cite{pro93}.
Additionally the solar ratios are given, too. The isotopic
chain of Ba has got unique s--process features. The abundance ratio of
the s--only isotopes $^{134}$Ba/$^{136}$Ba depends on a
branching at $^{134}$Cs which is sensitive to both s--process neutron
density and temperature. The isotope $^{138}$Ba controls the
average neutron
exposure $\tau_0$ in the s--process. The character of this dependence
shows the ratio
\begin{eqnarray}
^{138}{\rm Ba}/^{137}{\rm Ba}=
(\sigma_{138}+\tau_0^{-1})^{-1}/(\sigma_{137})^{-1} \quad ,
\end{eqnarray}
where the
MAC cross section $\sigma_{138}$ is comparable with $\tau_0^{-1}$.
A necessary condition that the exposure sensitivity of the
successive Ba--isotopes 136, 137,
and 138 on the unique synthesis path can be studied is the fact that
$^{138}$Ba/$^{136}$Ba is
significantly less than $(\sigma_{138})^{-1}/(\sigma_{136})^{-1}$,
i.e.,
there is a strong deviation from the anticorrelation between
abundance and  MAC cross section (compare e.g., the
s--process $^{138}$Ba/$^{136}$Ba ratios   
of Tab.\,\ref{t2a} with $(\sigma_{138})^{-1}/(\sigma_{136})^{-1}$ = 13.26
at
$kT$=8\,keV). 
In the SP-- and DP--model calculations the exposure $\tau_0$ has been
adjusted to reproduce the $^{138}$Ba/$^{136}$Ba ratio.
The s--Ba composition seems to be different from the
solar s--process (in brackets in Tab.\,\ref{t2a}) because the ratio
$^{138}$Ba/$^{136}$Ba is
different and consequently also the average exposure
(Table\,\ref{t2a}).

The support of the Volkswagen--Stiftung (Az: I/72286), the Deutsche
Forschungsgemeinschaft (DFG) (project Mo739) and the Fonds zur
F\"orderung der wissenschaftlichen Forschung (FWF S7307--AST) is
gratefully acknowledged.
TR is supported by an APART fellowship of the Austrian Academy of
Sciences.
\sectionnonum{References}


\begin{thebibliography}{9}
%
\bibitem{ohu96}
H. Oberhummer, H. Herndl, T. Rauscher, and H. Beer, Surveys in Geophys.,
{\bf 17} (1996) 665
%
\bibitem{bee91}
H. Beer, G. Rupp, F. Vo{\ss}, and F. K\"appeler, Ap. J. {\bf 379} (1991) 420
%
\bibitem{bee92}
H. Beer, M. Wiescher, F. K\"appeler, J. G\"orres, and P. E. Koehler,
Ap. J. {\bf 387} (1992) 258
%
\bibitem{bee95}
H. Beer, P. V. Sedyshev, Yu. P. Popov, W. Balogh, H. Herndl, and H.
Oberhummer, Phys. Rev. {\bf C52} (1995) 3442
%
\bibitem{bee96}
H. Beer, C. Coceva, P. V. Sedyshev, Yu. P. Popov, H. Herndl, R. Hofinger,
P. Mohr, and H. Oberhummer, Phys. Rev. {\bf C54} (1996) 2014
%
\bibitem{meis96a}
J. Mei{\ss}ner, H. Schatz, J. G\"orres, H. Herndl, M. Wiescher, H. Beer,
and F. K\"appeler, Phys. Rev. {\bf C53} (1996) 459
%
\bibitem{meis96b}
J. Mei{\ss}ner, H. Schatz, H. Herndl, M. Wiescher, H. Beer, and F.
K\"appeler, Phys. Rev. {\bf C53} (1996) 977
%
\bibitem{moh96}
P. Mohr, H. Oberhummer, and H. Beer, Proc. of the Int. Symposium on
Capture Gamma--Ray Spectroscopy and Related Topics, ed. G. Molnar, 
Budapest 1996, in print
%
\bibitem{moh97}
P. Mohr, H. Oberhummer, H. Beer, W. Rochow, V. Kn\"olle, G. Staudt,
P. V. Sedyshev, and Yu. P. Popov, Phys. Rev. {\bf C}, submitted
%
\bibitem{bcm97}
    H. Beer, F. Corvi, and P. Mutti, Ap. J. {\bf 474} (1997) 843
%
\bibitem{bee80}
H. Beer and F. K\"appeler, Phys. Rev. {\bf C21} (1980) 534
%
\bibitem{be94} H.\,Beer, G.\,Rupp, G.\,Walter, F.\,Vo{\ss}, and F.\,K\"appeler,
Nucl. Instr. Methods {\bf A337} (1994) 492
%
\bibitem{rat88} W.\,Ratynski and F.\,K\"appeler, Phys.\,Rev. {\bf C37}
(1988) 595; and R.\,L.\,Macklin, private communication
%
\bibitem{hop94}
P. Hoppe, S. Amari, E. Zinner, T. Ireland, and R. S. Lewis,
Ap. J. {\bf 430} (1994) 870
%
\bibitem{kim87}
K. H. Kim, M. H. Park, and B. T. Kim, Phys. Rev. {\bf C35} (1987) 363
%
\bibitem{kaep85}
F. K\"appeler, G. Walter, and G. J. Mathews, Ap. J. {\bf 291} (1985) 319
%
\bibitem{bt97}
H. Beer and T. Rauscher, Ap. J., submitted
%
\bibitem{tat87}
F.-K. Thielemann, M. Arnould, and J. W. Truran, in Adv. in Nucl.
Astrophys., eds. E. Vangioni-Flam et al. (Gif-sur-Yvette: Editions
Fronti\`eres 1987), p. 525
%
\bibitem{allen71} B. J. Allen, J. H. Gibbons,
and R. L. Macklin, Adv.\ Nucl.\ Phys.,\ {\bf 375} (1971) 823
%
\bibitem{Holmes76}
J. A. Holmes, S. E., Woosley, W. A. Fowler, and B. A. Zimmerman,
At.\ Data Nucl.\ Data Tables {\bf 18} (1976) 306
%
\bibitem{harris81}
M. J. Harris 1981, Astrophys. Space Sci. {\bf 77} (1981) 357
%
\bibitem{Cow91}J. J. Cowan, F.-K. Thielemann, and J. W. Truran,
 Phys.\ Rep.,\ {\bf 208} (1991) 267
%
\bibitem{tak87}
    K. Takahashi and K. Yokoi, Atomic Data Nucl. Data Tables
    {\bf 36} (1987) 375
%
\bibitem{cor91}
F. Corvi, G. Fioni, F. Gasperini, and P. B. Smith, Nucl. Sci. Eng.
{\bf 107} (1991) 272
%
\bibitem{mut96}
    P. Mutti, F. Corvi, K. Athanassopulos, and H. Beer,
    in {\it Nuclei in the Cosmos}, Notre Dame, USA,
    20.--27. June, 1996, ed. M. Wiescher,
    Nucl. Phys. A (Suppl.), in print
%
\bibitem{mut97}
P. Mutti, F. Corvi, K. Athanassopulos, and H. Beer, Int. Conf. on
{\it Nuclear Data for Science and Technology}, Trieste, Italy, 19.--24.
May, 1997
%
\bibitem{vwk95}
F. Voss, K. Wisshak, and F. K\"appeler, Phys. Rev. {\bf C52} (1995) 1102
%
\bibitem{ksw96}
P. E. Koehler, R. R. Spencer, R. R. Winters, K. H. Guber, J. A. Harvey, N.
W. Hill, and M. S. Smith, Phys. Rev. {\bf C54} (1996) 1463
%
\bibitem{mug84}
S. F. Mughabghab, Neutron Cross Sections, Vol 1, Part B (Academic Press,
INC., Orlando 1984)
%
\bibitem{pabe93}
H. Palme and H. Beer, in Landolt B\"ornstein, New Series VI, Astronomy
and Astrophysics, Subvolume 3a, (Berlin: Springer 1993), p. 196
%
\bibitem{pro93}
    C.A. Prombo, F.A. Podosek, S. Amari, and R.S. Lewis, Ap. J. {\bf 410}
(1990) 393
%



\end{thebibliography}
\end{document}